\documentclass[showpacs,preprintnumbers,prd,nofootinbib,floats,amssymb,floatfix]{revtex4}
\usepackage{amsmath}
\usepackage{amsfonts}
\usepackage{graphicx}
\usepackage{hyperref}
\usepackage{amsmath}
\usepackage{amstext}
 
\begin{document}
\title{Hamiltonian dynamics:   four dimensional  BF-like theories   with a compact dimension }  
    \author{ Alberto Escalante } \email{aescalan@ifuap.buap.mx}
    \author{Moises Zarate Reyes} \email{mzarate@ifuap.buap.mx}
 \affiliation{Instituto de F{\'i}sica Luis Rivera Terrazas, Benem\'erita Universidad Aut\'onoma de Puebla, (IFUAP).
   Apartado postal      J-48 72570 Puebla. Pue., M\'exico}   
\begin{abstract} 
A detailed Dirac's canonical analysis for a topological four dimensional $BF$-like theory with a compact dimension is developed.  By performing the compactification process  we find out   the relevant symmetries of the theory, namely,   the full structure of the constraints and   the extended action. We show that  the extended  Hamiltonian is a linear combination of first class constraints, which  means that the general covariance of the theory is not affected by the compactification process. Furthermore, in order to 
 carry out  the correct counting of physical  degrees of freedom,  we show that  must be taken into account  reducibility conditions among the first class constraints  associated to the excited KK modes. Moreover, we perform the Hamiltonian 
 analysis of Maxwell theory written as a $BF$-like  theory with a compact dimension, we analyze the   constraints of the theory and  we calculate the fundamental Dirac's brackets, finally the results obtained are compared with those  found in the literature.  
\end{abstract}
\date{\today}
\preprint{}
\maketitle
\section{INTRODUCTION}
Models that involve extra dimensions have introduced completely new ways of
 looking up on old problems in theoretical physics; the possible existence of a dimension 
 extra beyond the fourth dimension was considered around 1920's, when   Kaluza and Klein (KK) tried to unify electromagnetism 
 with Einstein's gravity by proposing a theory in 5D  where the fifth dimension is a circle  $S^1$ of radius $R$, and 
 the gauge field   is   contained in  the extra component of the  metric tensor  \cite{1}.  Nowadays,   the study of models involving extra dimensions  have an   important activity  in order to explain and solve 
 some  fundamental  problems found in theoretical physics,  such as,  the problem of  mass hierarchy,   the explanation of dark energy,  dark matter and inflation  \cite{1a}. Moreover, extra dimensions become also  important  in theories of  grand unification trying  of incorporating   gravity and gauge interactions consistently. In this respect, it is well known that extra dimensions have a  fundamental role in the developing of  string theory, since all versions of the theory are natural and consistently formulated only in a spacetime of more than four dimensions \cite{2, 3}. For some time, however, it was conventional to assume that in string  theory such extra dimensions were compactified to complex manifolds of small sizes   about the order of the  Planck length, 
$\ell_P\sim ~ 10^{-33}$~cm \cite{3, 6},  or  they  could be even  of lower size independently of the Plank Length \cite{6a, 6b, 6d}; in this respect, the compactification process is a crucial step in the construction of models with extra dimensions \cite{5}.  \\
On the other hand, there are phenomenological and theoretical motivations to quantize a gauge  theory in extra dimensions, for instance, if  there exist extra dimensions, then  their  effects  could be tested in the actual  LHC collider, and  in the International Linear Collider \cite{LHC-ILC}.\\
We can find  several works involving extra dimensions, for instance, in \cite{3, 5, 6} is developed   the canonical  analysis of Maxwell theory in five dimensions with a compact dimension,  after  performing  the compactification and  fixing the gauge parameters,    the final  theory describes to Maxwell theory plus a tower of KK excitations corresponding to massive Proca fields. Furthermore, in the context of Yang-Mills (YM) theories, in \cite{7} it has been carry out the canonical analysis of a 5D YM theory with a compact dimension;  in that work were obtained  different scenarios for the 4D effective action obtained after the compactification;   if the gauge
parameters propagate in the bulk, then  the excited KK modes are gauge fields, and  they are matter vector fields provided that  those  parameters are confined in the 3-brane. \\
On the other hand,   the study of alternative models  describing Maxwell and YM theories expressed  as the  coupling of topological theories have attracted attention recently because of its close relation with gravity. In fact, the study of   topological actions has been motived
in several contexts of theoretical physics given their interesting relation  with physical theories. One example of this is
the well-known MacDowell-Maunsouri formulation of gravity. In
this formulation, breaking the $SO(5)$ symmetry of a $BF$-theory for
$SO(5)$ group down to $SO(4)$  we can obtain the Palatini action plus
the sum of second Chern and Euler topological invariants. Due to
these topological classes have trivial local variations that do not
contribute classically to the dynamics, we thus obtain essentially
general relativity \cite{7a, 7b}.  Furthermore, in \cite{8, 9}, an analysis of specific limits in the gauge coupling of topological theories yielding a pure YM dynamics in four and three dimensions has been reported. In this respect, in the four-dimensional  case, nonperturbative topological configurations of the gauge fields are defined as having an important role in realistic theories, e.g. quantum chromodynamics. Moreover, the 3D case is analyzed at the Lagrangian level, and the action becomes  the coupling of $BF$-like terms  in order to generalize the quantum dynamics of YM and thus proving  possible extensions to the quantum dynamics of 3D gravity \cite{8}.   \\
Because of the ideas expressed above, in  this paper we analyze  a  four dimensional $BF$-like theory and the  Maxwell theory written as a $BF$-like theory with a compact dimension. First,  we perform the analysis for the $BF$ term;  in this case  we are interested  in  knowing  the symmetries of a topological theory defined in four  dimensions with  a compact dimension.  We shall show that in order to obtain the correct  counting of physical degrees of freedom, we  must  take  into account reducibility conditions   among the first class constraints of the KK excitations;  hence, in this paper we present  the study of a model with  reducibility conditions  in the KK modes. Finally, we perform the Hamiltonian analysis of  Maxwell theory written as the coupling of  a $BF$-like terms  with a compact dimension,  and we  compare our results with those found in the literature. In addition, we have added as appendix the fundamental Dirac's brackets of the theories under study, thus we develop  the first steps for studying the quantisation aspects.  \\
\setcounter{equation}{0} \label{c2}
\section{Hamiltonian dynamics of a BF-like topological theory with a compact dimension}
In the following lines, we shall study the Hamiltonian dynamics for a four dimensional  $BF$-like topological theory with a compact dimension;  then  we develop the canonical analysis of a four dimensional Maxwell theory written as a $BF$-like theory  with a compact dimension. \\ Let us start  with  the following  action  reported   in  Sundermeyer's  book  \cite{sund}  ( also   Yang-Mills theory is  written as a $BF$-like theory  in that book) defined in four dimensions  
\begin{equation}
S _1\left[ A, 	\textbf{B} \right]= \int d^{4}x \left\{ \frac{1}{4} B{^{M N}}B_{M N} - 
\frac{1}{2}B^{M N} \left( \partial_{M} A_{N} -\partial_{N} A_{M}\right)  \right\},
\label{ac1}
\end{equation}
where $B^{MN}=-B^{NM}$.    The equations of motion obtained from (\ref{ac1}) are given by  
\begin{equation}
\partial^{M}B_{MN}=0,
\label{max1}
\end{equation}
\begin{equation}
B_{MN}=\partial_{M}A_{N}-\partial_{N}A_{M},
\label{max2}
\end{equation}
by taking into account (\ref{max2}) in  (\ref{max1}), we obtain the Maxwell's free field equations. A pure Dirac's  analysis of the action (\ref{ac1}) has been reported in \cite{ 10},  where it  was showed that the action can be split   in
two terms lacking  physical degrees of freedom, the complete action, however, does have physical degrees of freedom, the Maxwellian  degrees of freedom. The studio of action (\ref{ac1}) with a compact dimension  becomes important because   could expose some information among the topological sector  given in the second  term  on the right hand side  of  (\ref{ac1}) (the $BF$ term),  and the dynamical sector given in the full action.  Hence, before continuing  with the analysis of (\ref{ac1}), we first analyce  the  following action  given by
\begin{equation}
S_{2}\left[ A, 	\textbf{B} \right]= \int d^{4}x  \left\{B^{M N} \left( \partial_{M} A_{N} -\partial_{N} A_{M}\right)  \right\}. 
\label{eq4a}
\end{equation}
The action $S_2$ is a topological theory, and its study in the context of extra dimensions   become relevant.  We need to remember that  topological field theories are characterized by being devoid of local degrees of freedom. That is, the theories are susceptible only to global degrees of freedom associated with non-trivial topologies of the manifold in which they are defined and topologies of the gauge bundle, thus the next question  arises;   it  is affected the topological nature  of $S_2$ because of the  compactification  process?.  Moreover, in order to carry out the  counting of  physical  degrees of freedom of (\ref{eq4a}) without a compact dimension,   we  must  take   into account  reducibility conditions among the constraints \cite{10, 12}, hence,      it is interesting to investigate if  reducibility constraints are still present after performing  the compactification process.  In fact, it has not been reported in the  literature  and  the Hamiltonian analysis of theories with reducibility conditions among the constraints  in the context of extra dimensions has not been performed,  and  we shall answer these questions along  this paper. \\ 
For simplicity we shall work with a four dimensional action,  then we will perform the compactification process in order to obtain a  three dimensional effective Lagrangian. It is straightforward  perform  the extension of our results to  dimensions higher than four.  The notation that we will use along the paper  is the following: the capital latin indices $M, N$ run over $0,1,2,3$ here $3$ label the compact dimension and these indices can be raised and lowered by the four-dimensional Minkowski metric $\eta_{M N}= (-1,1,1,1)$; $z$ will represent the coordinate in the compact dimension and $\mu, \nu=0,1,2$ are spacetime indices, $x^\mu$   the coordinates that label the points for the three-dimensional manifold $M_3$; furthermore we will suppose  that  the compact dimension is a $S^1/\mathbf{Z_2}$ orbifold whose radius is $R$; then any dynamical variable defined on  $M_{3}\times S^1/\mathbf{Z_2}$ can be expanded in terms of the complete set of harmonics  \cite{3, 6, 7, Muck:2002af}
\begin{eqnarray}\label{series}
B^{3\mu}(x,z)& = &\frac{1}{\sqrt{\pi R}}\sum_{n=1}^{\mathcal{\infty}}B^{3\mu}_{(n)}(x)\sin \left(\frac{n z}{R} \right),\nonumber\\
B^{\mu \nu}(x,z)& = &\frac{1}{\sqrt{2\pi R}}B^{\mu \nu}_{(0)}(x) + \frac{1}{\sqrt{\pi R}}\sum_{n=1}^{\mathcal{\infty}}B^{\mu \nu}_{(n)}(x)\cos \left(\frac{n z}{R} \right),\nonumber\\
A_{3}(x,z)& = &\frac{1}{\sqrt{\pi R}}\sum_{n=1}^{\mathcal{\infty}}A^{(n)}_{3}(x)\sin \left(\frac{n z}{R} \right),\nonumber\\
A_{\mu}(x,z)& = &\frac{1}{\sqrt{2\pi R}}A^{(0)}_{\mu}(x) + \frac{1}{\sqrt{\pi R}}\sum_{n=1}^{\mathcal{\infty}}A^{(n)}_{\mu}(x)\cos \left(\frac{n z}{R} \right).
\end{eqnarray}
The dynamical variables of the  theory  are given by  $A_{i}^{(0)}, A_{0}^{(0)}, B^{0i}_{(0)}, B^{i j}_{(0)}, A_{3}^{(n)}, A_{i}^{(n)}, A_{0}^{(n)}, B^{0 3}_{(n)}, B^{i3}_{(n)}, B^{0i}_{(n)}, B^{i j}_{(n)}$, with $ i,j=1,2.$\\
Let us perform the Hamiltonian analysis of the topological term  given by  $S_2$
\begin{equation}
S_{2}\left[ A, 	\textbf{B} \right]= \int d^{3}x \int_{0}^{2\pi R} dz \left\{B^{M N} \left( \partial_{M} A_{N} -\partial_{N} A_{M}\right)  \right\},
\label{eq3}
\end{equation}
first, we start the  analysis by performing the  3+1 decomposition and  we  use  explicitly the expansions  given in (\ref{series}); then we perform the  compactification  process on a $S^1/\mathbf{Z_2}$ orbifold,   obtaining the following  effective  Lagrangian    
\begin{equation}
\begin{aligned}
{\mathcal{L}_{2}}&=2B^{0i}_{(0)}\dot{A}^{(0)}_{i}
+2A_{0}^{(0)}\partial_{i}B^{0i}_{(0)}+B^{ij}_{(0)}F^{(0)}_{ij}+\sum_{n=1}^{\mathcal{\infty}}\Bigg[2A^{(n)}_{0}\partial_{i}B^{0 i}_{(n)}
+2B^{0i}_{(n)}\dot{A}^{(n)}_{i}+2B^{i3}_{(n)}\left(\partial_{i}A^{(n)}_{3}+\frac{n}{R}A^{(n)}_{i}\right)\\
&+2B^{0 3}_{(n)}\left(\partial_{0}A^{(n)}_{3}+\frac{n}{R}A^{(n)}_{0}\right)+B^{ij}_{(n)}F^{(n)}_{ij}\Bigg],\\
\end{aligned}
\label{eq5}
\end{equation}
where $F^{(m)}_{ij}=\partial_{i} A^{(m)}_{j} -\partial_{j} A^{(m)}_{i}$.  The first three terms on the left hand side  are called the zero modes and the theory describes a topological theory \cite{10, 12, 11},  the following terms correspond to a  KK tower;  in fact, both $B^{\alpha \beta}_{(n)}$ and $ A^{(n)}_{\alpha}$ are called Kaluza-Klein (KK) modes. 
In the following we shall suppose that the number of KK modes is given by $k$, taking the limit $k \rightarrow \infty$ at the end of the calculations. \\
The theory under study is a singular system, since  it is easy to observe that  the Hessian is a $10k - 4 \times 10k - 4$ matrix, it has determinant  equal to zero;  hence,  the Hamiltonian formalism calls for the definition of the momenta 
 $\left(\Pi^{(n)}_{MN}, \Pi^{M}_{(n)} \right)$ canonically conjugate to $\left(A^{(n)}_{M}, B^{M N}_{(n)}\right)$, 
\begin{equation}
\Pi^{M}_{(n)}= \frac{\delta L_{2}}{\delta \left(\partial_{0}A^{(n)}_{M}\right)}\;, \quad  \Pi^{(n)}_{M N}=\frac{\delta L_{2}}{\delta \left(\partial_{0}B^{M N}_{(n)}\right)}\;\;,
\label{mom}
\end{equation}
here, $n=1, 2, 3,.., k-1$. We commented that the determinant of the Hessian vanishes, we also can observe  that the rank  of the Hessian  is zero, so  we expect $10k-4$ primary constraints; from the definition of the momenta (\ref{mom}),  we identify the following  primary constraints: 
\begin{equation}
\begin{array}{ll}
\text{zero-modes}&\text{$k$-modes}\\
\phi^{(0)}_{0j} \equiv \Pi^{(0)}_{0j} \approx 0, & \phi^{(n)}_{03} \equiv \Pi^{(n)}_{03} \approx 0, \\ 
\phi^{(0)}_{ij} \equiv  \Pi^{(0)}_{ij} \approx 0, & \phi^{(n)}_{i3} \equiv \Pi^{(n)}_{i3} \approx 0,\\ 
\phi^{i}_{(0)} \equiv \Pi^{i}_{(0)}-2B^{0i}_{(0)} \approx  0, & \phi^{(n)}_{0i} \equiv \Pi^{(n)}_{0i} \approx 0,\\
\phi^{0}_{(0)}\equiv \Pi^{0}_{(0)} \approx  0, & \phi^{(n)}_{0i} \equiv \Pi^{(n)}_{0i} \approx 0,\\
    &\phi^{3}_{(n)} \equiv \Pi^{3}_{(n)}-2B^{03}_{(n)} \approx  0,\\
    &\phi^{i}_{(n)} \equiv \Pi^{i}_{(n)}-2B^{0i}_{(n)} \approx  0,\\
    &\phi^{0}_{(n)} \equiv \Pi^{0}_{(n)} \approx  0.
\end{array}\label{eq6}
\end{equation}
Furthermore, the canonical Hamiltonian is given by 
\begin{equation}
H_{c}=\int d^{2}x\Bigg(-A^{(0)}_{0}\partial_{i}\Pi^{i}_{(0)}-B^{ij}_{(0)}F^{(0)}_{ij}+\sum_{n=1}^{\mathcal{\infty}}\Big[-2B^{i3}_{(n)}\left(\partial_{i}A^{(n)}_{3}+\frac{n}{R}A^{(n)}_{i}\right)-A^{(n)}_{0}\left(\partial_{i}\Pi^{i}_{(n)}+\frac{n}{R}\Pi^{3}_{(n)}\right)
- \nonumber \\ B^{ij}_{(n)}F^{(n)}_{ij}\Big]\Bigg) \nonumber
\label{eq7}
\end{equation} 
thus by using the primary constraints   (\ref{eq6}),   we define  the primary Hamiltonian given by 
\begin{eqnarray}
H_{P}&= &H_{c} + \int dx^2 \Bigg[\lambda_{(0)}^{0j}\phi^{(0)}_{0j}+\lambda_{(0)}^{ij}\phi^{(0)}_{ij}
+\lambda^{(0)}_{i}\phi^{i}_{(0)}+\lambda^{(0)}_{0}\phi^{0}_{(0)}+\sum_{n=1}^{\mathcal{\infty}}\Bigg(\lambda^{03}_{(n)} \phi_{03}^{(n)} + \lambda^{j3}_{(n)} \phi_{j3}^{(n)}+\lambda^{(n)}_{3}\phi^{3}_{(n)} +\lambda_{(n)}^{0j}\phi^{(n)}_{0j}\nonumber\\
& &+\lambda_{(n)}^{ij}\phi^{(n)}_{ij}+\lambda^{(n)}_{i}\phi^{i}_{(n)}+\lambda^{(n)}_{0}\phi^{0}_{(n)}\Bigg)\Bigg],
\label{eq8} 
\end{eqnarray}
where $\lambda^{03}_{(n)}$,  $\lambda^{j3}_{(n)}, \lambda^{(n)}_{3}, \lambda_{(n)}^{0j}, \lambda_{(n)}^{ij}, \lambda^{(n)}_{i}, \lambda^{(n)}_{0}$ and $\lambda_{(0)}^{0j}, \lambda_{(0)}^{ij}, \lambda^{(0)}_{i}, \lambda^{(0)}_{0}$  are Lagrange multipliers enforcing the  constraints.
The non-vanishing fundamental Poisson brackets for the theory under study are given by 
\begin{eqnarray}
\{A^{(m)}_{M}(x^0, x), \Pi^{N}_{(n)} (x^0, y) \} &=& \delta{^{M}}_{N}\delta{^{m}}_{n}\delta^2(x-y), \nonumber \\ 
\{B^{M N}_{(m)}(x^0, x), \Pi^{(n)}_{IJ}(x^0,y) \} &=& \frac{1}{2}\delta{^{m}}_{n}\left( \delta{^{M}}_{I} \delta{^{N}}_{J} -\delta{^{N}}_{I} \delta{^{M}}_{J}\right) \delta^2(x-y).
\label{eq9}
\end{eqnarray}
Let us now analyze  if  secondary constraints arise from the consistency conditions over the primary constraints. For this aim,  we construct  the $(10k-4)\times(10k-4)$ matrix formed by the Poisson brackets among  the primary constraints;  the non-vanishing  Poisson brackets between primary constraints are given by
\begin{eqnarray*}
\{ \phi^{(0)}_{0i} (x), \phi^{j}_{(0)}(y) \}&=&\delta{^{j}}_{i}\delta^2(x-y),\nonumber\\
\{ \phi^{(m)}_{03} (x), \phi^{3}_{(n)}(y) \}&=&\delta{^{m}}_{n}\delta^2(x-y),\nonumber\\   
\{ \phi^{(m)}_{0i} (x), \phi^{j}_{(n)}(y) \}&=&\delta{^{m}}_{n}\delta{^{j}}_{i}\delta^2(x-y),\nonumber\\
\label{eq11a}
\end{eqnarray*}
that matrix has rank=$6k-2$ and  $4k-2$ null vectors. From consistency and by using the null vectors,  we find the following $4k-2$ secondary constraints 
\begin{eqnarray}
\dot{\phi}^{0}_{(0)}(x)= \{\phi^{0}_{(0)}(x), {H}_{P} \} \approx 0 \quad &\Rightarrow& \quad  \psi_{(0)}=\partial_{k}\Pi^{k}_{(0)},\approx 0.\nonumber\\ 
\dot{\phi}^{(0)}_{ij}(x)= \{\phi^{(0)}_{ij}(x), {H}_{P} \} \approx 0 \quad &\Rightarrow& \quad  \psi^{(0)}_{ij}=F^{(0)}_{ij}\approx 0,\nonumber\\
\dot{\phi}^{(m)}_{k3}(x)= \{\phi^{(m)}_{k3}(x), {H}_{P} \} \approx 0 \quad & \Rightarrow& \quad \psi^{(m)}_{k3}=\partial_{k}A^{(m)}_{3}+\frac{m}{R}A^{(m)}_{k}\approx 0, \nonumber \\
\dot{\phi}^{0}_{(m)}(x)= \{\phi^{0}_{(m)}(x), {H}_{P} \} \approx 0 \quad &\Rightarrow& \quad  \psi^{3}_{(m)}=\partial_{k}\Pi^{k}_{(m)}+\frac{m}{R}\Pi^{}_{(m)} \approx 0,\nonumber\\ 
\dot{\phi}^{(m)}_{ij}(x)= \{\phi^{(m)}_{ij}(x), {H}_{P} \} \approx 0 \quad &\Rightarrow& \quad  \psi^{(m)}_{ij}=F^{(m)}_{ij}\approx 0;
\label{eq12}
\end{eqnarray}
 and the rank allows us to fix the following $6k-2$ Lagrange multipliers
 \begin{equation}
\begin{array}{ll}
\text{zero-modes}&\text{$k$-modes}\\
\lambda^{0j}_{(0)}=-2\partial_{i}B^{ij}_{(0)}, & \lambda^{03}_{(n)}=-2\partial_{i}B^{i3}_{(n)}, \\ 
\lambda^{(0)}_{i}=0,& \lambda^{(n)}_{3}=0,\\ 
     & \lambda_{(m)}^{0k}=-2\partial_{i}B^{ik}_{(m)}+\frac{2m}{R}B^{k3}_{(m)},\\
     & \lambda^{(m)}_{i}=0, \\
\end{array}     
\label{eq12}
\end{equation}
 for this theory there are not third constraints. Hence, this completes Dirac's consistency procedure for
finding the complete set of constraints; the set  of constraints primary and secondary  obtained are given by 
\begin{equation}
\begin{array}{ll}
\text{zero-modes}&\text{$k$-modes}\\
\phi^{(0)}_{0j}\equiv \Pi^{(0)}_{0j}\approx 0,&\phi^{(n)}_{0 3}\equiv \Pi^{(n)}_{0 3}\approx 0,\\
\phi^{(0)}_{ij}\equiv \Pi^{(0)}_{ij}\approx 0,&\phi^{(n)}_{i3}\equiv \Pi^{(n)}_{i 3}\approx 0,\\
\phi^{i}_{(0)}\equiv \Pi^{i}_{(0)}-2B^{0 i}_{(0)}\approx 0,&\phi^{(n)}_{0i}\equiv \Pi^{(n)}_{0 i}\approx 0,\\
\phi^{0}_{(0)}\equiv \Pi^{0}_{(0)}\approx 0,& \phi^{(n)}_{ij}\equiv \Pi^{(n)}_{i j}\approx 0,\\
\psi^{0}_{(0)}\equiv \partial_{k}\Pi^{k}_{(0)}\approx 0,& \phi^{3}_{(n)}\equiv \Pi^{3}_{(n)}-2B^{0 3}_{(n)}\approx 0,\\
\psi^{(0)}_{ij}\equiv F^{(0)}_{ij}\approx 0,& \phi^{i}_{(n)}\equiv \Pi^{i}_{(n)}-2B^{0 i}_{(n)}\approx 0,\\
        & \phi^{0}_{(n)}\equiv \Pi^{0}_{(n)}\approx 0, \\
        & \psi^{(n)}_{k3}\equiv \partial_{k}A^{(n)}_{3}+\frac{n}{R}A^{(n)}_{k}\approx 0,  \\
        & \psi^{3}_{(n)}\equiv \partial_{k}\Pi^{k}_{(n)}+\frac{n}{R}\Pi^{3}_{(n)}\approx 0,  \\
        & \psi^{(n)}_{ij}\equiv F^{(n)}_{ij}\approx 0.\\
\end{array}\label{eq11a}
\end{equation}
Once identified all the constraints as  primary, secondary etc., we need to know  which ones correspond to first and second class. For this  purpose we will construct  the matrix formed by the Poisson brackets among  the primary and secondary constraints; in order to achieve  this aim,  the non-zero  Poisson brackets  among   primary and secondary constraints  are given by  
\begin{eqnarray}\label{eq13-1}
\{ \phi^{(0)}_{0i} (x), \phi^{(0)j}(y) \}&=&\delta{^{j}}_{i}\delta^2(x-y),\nonumber\\ 
\{ \phi^{j}_{(0)}(x), \psi^{(0)}_{ls}(y) \}& =& -\left( \delta{^{j}}_{s}\partial^{y}_{l}-\delta{^{j}}_{l}\partial^{y}_{s}\right) \delta^2(x-y),\nonumber \\
\{ \phi^{(m)}_{03} (x), \phi^{3}_{(n)}(y) \}&=&\delta{^{m}}_{n}\delta^2(x-y),\nonumber\\   
\{ \phi^{(m)}_{0i} (x), \phi^{j}_{(n)}(y) \}&=&\delta{^{m}}_{n}\delta{^{i}}_{j}\delta^2(x-y),\nonumber\\
\{ \phi^{3}_{(m)}(x),\psi^{(n)}_{k3}(y) \}& =&-\delta{^{m}}_{n}\partial^{y}_{k}\delta^2(x-y),\nonumber \\
\{ \phi^{j}_{(m)}(x), \psi^{(n)}_{k3}(y) \}& =&-\frac{n}{R}\delta{^{m}}_{n}\delta{^{k}}_{j}\delta^2(x-y),\nonumber \\  
\{ \phi^{j}_{(m)}(x), \psi^{(n)}_{ls}(y) \}& =& -\delta{^{m}}_{n}\left( \delta{^{j}}_{s}\partial^{y}_{l}-\delta{^{j}}_{l}\partial^{y}_{s}\right) \delta^2(x-y).
\end{eqnarray}
That matrix has a rank $=6k-2$ and $8k-4$ null vectors,   thus,  by using the rank and the null vectors,  we find  the following 4  first class constraints  for the zero modes 
\begin{eqnarray}
\tilde{\gamma}^{(0)}_{ij} &=&F^{(0)}_{ij}-\partial_{i}\Pi^{(0)}_{0j}+\partial_{j}\Pi^{(0)}_{0i}\approx 0,\nonumber \\
\gamma_{(0)}&=&\partial_{i}\Pi^{i}_{(0)} \approx 0,\nonumber \\
\gamma^{(0)}_{ij}&=&\Pi^{(0)}_{ij} \approx 0,\nonumber \\
\gamma^{0}_{(0)}&=&\Pi^{0}_{(0)} \approx 0,
\label{eq15u}
\end{eqnarray}
and the following  4 second class constraints for the zero modes
\begin{eqnarray}
\chi^{(0)}_{0i}&=&\Pi^{(0)}_{0i} \approx 0,\nonumber\\
\chi^{i}_{(0)}&=&\Pi^{i}_{(0)}-2B^{0i}_{(0)}\approx 0.
\label{eq15}
\end{eqnarray}
Furthermore, we  identifying  the following $8k-8$ first class constraints for the KK-modes 
\begin{eqnarray} 
\tilde{\gamma}^{(m)}_{i3} &=&\partial_{i}A^{(m)}_{3}+\frac{m}{R}A^{(m)}_{i}-\partial_{i}\Pi^{(m)}_{03}-\frac{m}{R}\Pi^{(m)}_{0i} \approx 0,\nonumber \\
\tilde{\gamma}^{(m)}_{ij} &=&F^{(m)}_{ij}-\partial_{i}\Pi^{(m)}_{0j}+\partial_{j}\Pi^{(m)}_{0i} \approx 0,\nonumber \\
\gamma^{(m)}_{i3}&=&\Pi^{(m)}_{i3}\approx 0,\nonumber \\
\gamma^{(m)}_{ij}&=&\Pi^{(m)}_{ij} \approx 0,\nonumber \\
\gamma^{0}_{(m)}&=&\Pi^{0}_{(m)} \approx 0,\nonumber \\
\gamma_{(m)}&=&\partial_{i}\Pi^{i}_{(m)}+\frac{m}{R}\Pi^{3}_{(m)} \approx 0,
\label{eq14u}
\end{eqnarray}
and $6k-6$ second class constraints 
\begin{eqnarray}     
\chi^{(m)}_{03}&=&\Pi^{(m)}_{03} \approx 0,\nonumber \\
\chi^{3}_{(m)}&=&\Pi^{3}_{(m)}-2B^{03}_{(m)} \approx 0,\nonumber\\
\chi^{i}_{(m)}&=&\Pi^{i}_{(m)}-2B^{0i}_{(m)} \approx 0,\nonumber\\
\chi^{(m)}_{0i}&=&\Pi^{(m)}_{0i} \approx 0.
\label{eq14}
\end{eqnarray}
With all this information at hand, the counting of degrees of freedom is carry out as follows: there are $20k-8$  dynamical variables,  $8k-4$ first class constraints and $6k-2$ second class constraints, therefore the number of degrees of freedom is given by
\begin{equation}
G=\frac{1}{2}\left(20k- 8 -\left(2(8k-4)+6k-2\right)\right)=-(k-1),
\end{equation} 
this is an interesting fact, the counting of degrees of freedom is negative and this can not be correct.  It is important to comment, that in a four dimensional $BF$ theory without a compact dimension,   in order to carry out the correct counting  of physical degrees of freedom,  we must   take into account   reducibility  conditions among the first class constraints \cite{12, 11}.  Hence, if we observe the constraints found above,  we can see that the reducibility among the constraints is also  present;  however, there exist   reducibility conditions in the first class constraints of the  KK excitations and there are not   in the zero mode.  In fact,  it  can be showed that  the reducibility conditions are identified  by the following $k-1$  relations 
\begin{eqnarray}\label{reduc}
\partial_{i}\tilde{\gamma}^{(m)}_{j3}-\partial_{j}\tilde{\gamma}^{(m)}_{i3}-\frac{m}{R}\tilde{\gamma}^{(m)}_{ij}=0,
\end{eqnarray}
in this manner, the number of independent first class constraints are $(8k-4- k+1=7k-3)$; then, this implies that the number of physical degrees of freedom is 
\begin{equation}
G=\frac{1}{2}\left(20k- 8-\left(2(7k-3)+6k-2\right)\right)=0. 
\end{equation}
Therefore, the $BF$-like theory with a compact dimension  is still   topological one. It is important to comment that  if we perform the counting of physical degrees of freedom for the zero mode, then   we find that  it  is devoid of local degrees of freedom as expected; for the zero mode defined in three dimensions  there are not reducibility conditions.  All this information become relevant,  because after performing the compactification process  there are already reducibility conditions; we need to remember that the correct identification of the constraints  is a relevant step because they allows us identify observables  and constraints  are  the best guideline to perform the quantization;  similarly the reducibility conditions in the KK  modes must be taken into account in that process. \\
With all this  information, we can identify the extended action; thus,  we use  the first class constraints (\ref{eq14}), the second class constraints (\ref{eq15}),  the Lagrange multipliers (\ref{eq12}),  and we  find that  the extended action takes the form
\begin{equation}
\begin{split}
S_{E}\Big(Q_{K},P_{K},\lambda_{K}\Big)& =\int d^{3}x\Big[\dot{A}_{\nu}^{(0)}\Pi^{\nu}_{(0)}+ \dot{ B}^{\nu\mu}_{(0)}\Pi^{(0)}_{\nu\mu}-\mathcal{H}^{(0)}-\tilde{\alpha}^{ij}_{(0)}\tilde{\gamma}^{(0)}_{ij}
-\alpha^{(0)}\gamma_{(0)}-\alpha^{ij}_{(0)}\gamma^{(0)}_{ij}-\alpha^{(0)}_{0}\gamma^{0}_{(0)}\\
&-\lambda^{0i}_{(0)}\chi^{(0)}_{0i}-\lambda^{(0)}_{i}\chi^{i}_{(0)}+\sum_{n=1}^{k} \Big\{ \dot{A}_{N}^{(n)}\Pi^{N}_{(n)}+ \dot{ B}^{MN}_{(n)}\Pi^{(n)}_{MN}-\mathcal{H}^{(n)}-\alpha^{i3}_{(n)}\tilde{\gamma}^{(n)}_{i3}
-\alpha^{ij}_{(n)}\tilde{\gamma}^{(n)}_{ij}\\
&- \lambda^{i3}_{(n)}\gamma^{(n)}_{i3}
-\lambda^{ij}_{(n)}\gamma^{(n)}_{ij}
-\lambda^{(n)}_{0}\gamma^{0}_{(n)}-\alpha^{(n)}\gamma_{(n)}-\lambda^{(n)}_{i}\chi^{i}_{(n)}
-\lambda^{0i}_{(n)}\chi^{(n)}_{0i}-\lambda^{03}_{(n)}\chi^{(n)}_{03}\\
&-\lambda^{(n)}_{3}\chi^{3}_{(n)}\Big\}\Big],\\
\end{split}
\label{actbf}
\end{equation}
where we abbreviate with  $Q_{K}$ y $P_{K}$ all the dynamical variables and the generalized momenta;   $\lambda_{K}$ stand for  all  Lagrange multipliers associated with the first and second class constraints.
From  the extended action, it is possible to identify the extended Hamiltonian and is given by
\begin{equation}
\begin{split}
H_{ext}&=\int d^{2}x\Big[-A^{(0)}_{0}\gamma^{(0)}-B^{ij}_{(0)}\tilde{\gamma}^{(0)}_{i j}+\sum_{n=1}^{k}\Big[-A^{(n)}_{0}\gamma^{3}_{(n)}
-2B^{i3}_{(n)}\tilde{\gamma}^{(n)}_{i 3}-B^{ij}_{(n)}\tilde{\gamma}^{(n)}_{i j}\Big]+\alpha^{ij}_{(0)}\tilde{\gamma}^{(0)}_{ij}+\alpha^{(0)}\gamma_{(0)}\\
&+\lambda^{ij}_{(0)}\gamma^{(0)}_{ij}
+\alpha^{(0)}_{0}\gamma^{0}_{(0)}+\sum_{n=1}^{k} \Big\{\alpha^{i3}\tilde{\gamma}^{(n)}_{i3}+\alpha^{ij}_{(n)}\tilde{\gamma}^{(n)}_{ij}+ \lambda^{i3}_{(n)}\gamma^{(n)}_{i3}
+\lambda^{ij}_{(n)}\gamma^{(n)}_{ij}+\lambda^{(n)}_{0}\gamma^{0}_{(n)}
+\alpha^{(n)}\gamma_{(n)}\Big\}\Big],
\end{split}
\label{hextbf}
\end{equation}
we can observe  that this expression  is a linear combination of  constraints. In fact, they are  first class constraints of the zero mode and first class constrains of the $KK$-modes. It is well-known, that  for the action (\ref{eq3}) without compact dimensions,  its  extended Hamiltonian is a linear combination of first class constraints \cite{12, 11}, thus, we can notice that the general covariance of the theory is not affected by the compactification process.  Hence,  in order to perform a quantization of  the theory,  it is not possible to construct the Schrodinger equation because the
action of the Hamiltonian on physical states is annihilation. In Dirac's quantization of systems with general covariance, the restriction on  physical states is archived by demanding that the first class constraints in their quantum form must be satisfied;  thus in this paper we have all  tools for  studying  the quantization of the theory by means a canonical framework.\\
 By following with our analysis, we need to know the gauge transformations on the phase space. For this important step, we shall  define  the following gauge generator in terms of the first class constraints (\ref{eq14})  
\begin{eqnarray}
G&=&\int_\Sigma d^{2}x\Bigg[\varepsilon^{i3}_{(n)}\tilde{\gamma}^{(n)}_{i3}+\varepsilon^{ij}_{(n)}\tilde{\gamma}^{(n)}_{ij}
+\varepsilon^{(n)}_{0}\gamma_{(n)}+\dot{\varepsilon}^{i3}_{(n)}\gamma^{(n)}_{i3}+ \dot{\varepsilon}^{ij}_{(n)}\gamma^{(n)}_{ij}+\dot{\varepsilon}^{(n)}_{0}\gamma^{0}_{(n)}+\varepsilon^{ij}_{(0)}\tilde{\gamma}^{(0)}_{ij}+ \dot{\varepsilon}^{ij}_{(0)}\gamma^{(0)}_{ij}\nonumber\\
& &+\varepsilon^{(0)}_{0}\gamma^{0}_{(0)}
+\dot{\varepsilon}^{(0)}_{0}\gamma_{(0)}\Bigg], 
\label{eqgbf}
\end{eqnarray}
thus we obtain that the gauge transformations on the phase  space  are  given by
\begin{equation}
\begin{array}{ll}
\text{zero-mode}&\text{k-mode}\\
\delta A_{\mu}^{(0)}=-\partial_{\mu}\varepsilon^{(0)}_{0} ,&\delta A_{\mu}^{(n)} =-\partial_{\mu}\varepsilon^{(n)}_{0},\\ 
\delta B^{0i}_{(0)} =\partial_{k}\varepsilon^{k i}_{(0)},&\delta A_{3}^{(n)} =\frac{n}{R}\varepsilon^{(n)}_{0} ,\\ 
\delta B^{ij}_{(0)} =\partial_{0}\varepsilon ^{ij}_{(0)},&\delta B^{03}_{(n)}=\frac{1}{2}\partial_{i}\varepsilon^{i3}_{(n)},\\ 
\delta \Pi^{i}_{(0)} =\partial_{k}\varepsilon^{ki}_{(0)},&\delta B^{0i}_{(n)}=-\partial_{k}\varepsilon^{ik}_{(n)}-\frac{n}{2R}\varepsilon^{i3}_{(n)},\\ 
 \delta \Pi^{0}_{(0)} = 0, & \delta B^{i3}_{(n)} =\frac{1}{2}\partial_{0}\varepsilon ^{i 3}_{(n)}, \\ 
\delta \Pi^{M N}_{(0)} = 0,  & \delta B^{ij}_{(n)} =\partial_{0}\varepsilon ^{ij}_{(n)}, \\ 
  & \delta \Pi^{3}_{(n)}=-\partial_{i}\varepsilon^{i3}_{(n)},\\
  &\delta \Pi^{i}_{(n)} =\partial_{k}\varepsilon^{ki}_{(n)}-\frac{n}{R}\varepsilon^{i3}_{(n)},\\
  &\delta \Pi^{0}_{(n)} = 0,\\
  &\delta \Pi^{M N}_{(n)} = 0.
\end{array}\label{eqtrafis}
\end{equation}
We are able to notice that the fields $B^{MN}$ and $A_M$ are gauge fields; there are not degrees of freedom, thus, it  is not relevant  to  fix  the gauge parameters. In the following lines, we shall perform the Hamiltonian analysis of the action (\ref{ac1}) and we will find that the field $B^{MN}$ is not a gauge field anymore,  there are not reducibility conditions among the constraints, moreover, there exist  physical degrees of freedom and the fixing of the gauge parameters will allow us  to find  massive Proca fields and pseudo-Goldston bosons as  expected. Furthermore,  we have added in the   appendix B the  Dirac  brackets of the theory being an important step for studying the quantization. \\
\section{Hamiltonian analysis of the four-dimensional Maxwell theory written as a BF-like theory  with a compact dimension}
By following the steps developed above, we can perform the Hamiltonian analysis of (\ref{ac1}).  In this section we shall  resume the complete  analysis;   thus by  performing  the $3+1$ decomposition,  using the expansion of the fields  (\ref{series}),  and developing   the compactification  process on a $S^1/\mathbf{Z_2}$ orbifold,  we  obtain the following effective Lagrangian written as
\begin{equation}
{\mathcal{L}}=\frac{1}{4}B^{\mu \nu}_{(0)}B^{(0)}_{\mu \nu}-\frac{1}{2}B^{\mu \nu}_{(0)}F^{(0)}_{\mu \nu}+\sum_{n=1}^{\mathcal{\infty}}\Bigg[\frac{1}{2}B^{\nu 3}_{(n)}B^{(n)}_{\nu 3}-B^{\mu 3}_{(n)}\left(\partial_{\mu}A^{(n)}_{3}+\frac{n}{R}A^{(n)}_{\mu}\right)+\frac{1}{4}B^{\mu \nu}_{(n)}B^{(n)}_{\mu \nu}-\frac{1}{2}B^{\mu \nu}_{(n)}F^{(n)}_{\mu \nu}\Bigg].
\label{eqlaga}
\end{equation}
We are able to identify  the zero mode  given by $\frac{1}{4}B^{\mu \nu}_{(0)}B^{(0)}_{\mu \nu}-\frac{1}{2}B^{\mu \nu}_{(0)}F^{(0)}_{\mu \nu}$ and  the following terms are identified as the  KK excitations. We  have commented above, the action (\ref{eqlaga}) describes Maxwell theory in three dimensions (zero mode) plus a tower of KK-modes. The theory is singular, there  exists  the same number of dynamical variables  defined above. Hence,  after developing    a pure Dirac's analysis,  we find a set of $2k$ first class constraints given by 
\begin{eqnarray}
\gamma^{0}_{(0)}&=&\Pi^{0}_{(0)} \approx 0,  \nonumber \\
\gamma_{(0)}&=&\partial_{i}\Pi^{i}_{(0)} \approx 0,  \nonumber \\
\gamma^{0}_{(n)}&=&\Pi^{0}_{(n)} \approx 0, \nonumber \\
\gamma_{(n)}&=&\partial_{i}\Pi^{i}_{(n)}+\frac{n}{R}\Pi^{3}_{(n)}\approx 0, 
\label{eq27a}
\end{eqnarray}
and the following $12k-6$ second class constraints
\begin{eqnarray}
\chi^{(0)}_{0i}&=&\Pi^{(0)}_{0i} \approx 0,\nonumber \\
\chi^{(0)}_{ij}&=&\Pi^{(0)}_{ij} \approx 0,\nonumber \\
\chi^{j}_{(0)}&=&\Pi^{j}_{(0)}+B^{0j}_{(0)} \approx 0,\nonumber \\
\tilde{\chi}^{(0)}_{ij}&=&\frac{1}{2}\left(B^{(0)}_{ij}-F^{(0)}_{ij}\right) \approx 0,\nonumber \\
\chi^{(n)}_{03} &=&\Pi^{(n)}_{03} \approx 0,  \nonumber \\
\chi^{(n)}_{i3} &=&\Pi^{(n)}_{i3} \approx 0,  \nonumber \\
\chi^{(n)}_{0j} &=&\Pi^{(n)}_{0j} \approx 0,  \nonumber \\
\chi^{(n)}_{ij} &=&\Pi^{(n)}_{ij} \approx 0,  \nonumber \\
\chi^{3}_{(n)}&=&\Pi^{3}_{(n)}+B^{03}_{(n)} \approx 0, \nonumber \\
\chi^{i}_{(n)}&=&\Pi^{i}_{(n)}+B^{0i}_{(n)} \approx 0, \nonumber \\
\tilde{\chi}^{(n)}_{i3}&=&\frac{1}{2}\left(B^{(n)}_{i3}-\left(\partial_{i}A^{(n)}_{3}+\frac{n}{R}A^{(n)}_{i}\right)\right) \approx 0,  \nonumber \\
\tilde{\chi}^{(n)}_{ij}&=&\frac{1}{2}\left(B^{(n)}_{ij}-F^{(n)}_{ij}\right) \approx 0,  
\label{eq28a}
\end{eqnarray}
The identification of second class constraints, allows us to fix the following $12k-6$ Lagrange multipliers 
\begin{equation}
\begin{array}{ll}
\text{zero-modes}&\text{$k$-modes}\\
\lambda^{0k}_{(0)}=-4\partial_{i}B^{i k}_{(0)}+2\partial_{i}F^{i k}_{(0)}, & \lambda^{0 3}_{(n)}=-4\partial_{i}B^{i 3}_{(n)}+2\partial_{i}\left(\partial_{i}A^{(n)}_{3}+\frac{n}{R}A^{(n)}_{i}\right),\\ 
\lambda^{ij}_{(0)}=\partial^{i}\Pi^{j}_{(0)}-\partial^{j}\Pi^{i}_{(0)}, & \lambda^{i 3}_{(n)}= 2\big(\partial^{i}\Pi^{3}_{(n)}+\frac{n}{R}\Pi^{i}_{(n)}\big), \\ 
\lambda^{(0)}_{i}=0,& \lambda^{0 i}_{(n)}=-4\partial_{k}B^{k i}_{(n)}+2\partial_{k}F^{k i}_{(n)}+\frac{2n}{R}\left(\partial_{i}A^{(n)}_{3}+\frac{n}{R}A^{(n)}_{i}\right), \\ 
\beta^{ij}_{(0)}=B^{i j}_{(0)}-F^{i j}_{(0)}, & \lambda^{ij}_{(n)}=\partial^{i}\Pi^{j}_{(n)}-\partial^{j}\Pi^{i}_{(n)}, \\ 
     & \lambda^{(n)}_{3}=0, \\ 
     & \lambda^{(n)}_{i}=0, \\ 
     & \beta^{i3}_{(n)}=2\left(B^{i 3}_{(n)}-F^{i 3}_{(n)}\right),\\
     & \beta^{ij}_{(n)}=B^{i j}_{(n)}-F^{i j}_{(n)}.
\end{array}
\end{equation} 
By using all this  information it is possible to carryout the counting of degrees of freedom as follows; there are  $10k-4$ dynamical variables, $ 2k $ first class constraints 
and $12k-6$ second class constraints, thus 
\begin{eqnarray*}
G&=&\frac{1}{2}\left(20k-8-\left(2(2k)+12k-6\right)\right)\nonumber\\
&=& 2k-1.\nonumber
\end{eqnarray*} 
We observe that if $k=1$ we obtain one degree of freedom as  expected for Maxwell theory in three dimensions. \\
 By using the first class constraints (\ref{eq27a}),  the second class constraints (\ref{eq28a}), and the Lagrange multipliers   we find that the extended action takes the form
\begin{equation}
\begin{split}
S_{E}\Big(Q_{K},P_{K},\lambda_{K}\Big)&=\int\Big[\dot{A}_{\nu}^{(0)}\Pi^{\nu}_{(0)}+\dot{ B}^{(0)}_{\nu\mu}\Pi^{\nu\mu}_{(0)}-\mathcal{H}^{(0)}-\beta^{(0)}\gamma_{(0)}-\lambda^{(0)}_{0}\gamma^{0}_{(0)}
-\lambda^{(0)}_{i}\chi^{i}_{(0)}-\lambda^{ij}_{(0)}\chi^{(0)}_{ij}-\beta^{ij}_{(0)}\tilde{\chi}^{(0)}_{ij}\\
&+\sum_{n=1}^{\mathcal{N}} \Big\{ \dot{A}_{N}^{(n)}\Pi^{N}_{(n)}+ \dot{ B}^{(n)}_{MN}\Pi^{MN}_{(n)}-\mathcal{H}^{(n)}-\lambda^{(n)}_{0}\gamma^{0}_{(n)}- \beta^{(n)}\gamma_{(n)}- \lambda^{(n)}_{i}\chi^{i}_{(n)}-\lambda^{(n)}_{3}\chi^{3}_{(n)}\\
&-\lambda^{0 3}_{(n)}\chi^{(n)}_{0 3}-\lambda^{0i}_{(n)}\chi^{(n)}_{0i}-\lambda^{i3}_{(n)}\chi^{(n)}_{i3}-\lambda^{ij}_{(n)}\chi^{(n)}_{ij}-\beta^{i3}_{(n)}\tilde{\chi}^{(n)}_{i3}
-\beta^{ij}_{(n)}\tilde{\chi}^{(n)}_{ij}\Big\} \Big]dx^3,
\end{split} 
\label{actmax}
\end{equation}
where the corresponding extended Hamiltonian is given by
\begin{eqnarray}
H_{ext}=H+\int\Big[\beta_{(0)}\gamma^{(0)}+\lambda^{(0)}_{0}\gamma^{0}_{(0)}+\sum_{n=1}^{\mathcal{N}} \Big\{\lambda^{(n)}_{0}\gamma^{0}_{(n)}+\beta^{(n)}\gamma_{(n)}\Big\}\Big]dx^3,
\label{hamax}
\end{eqnarray}
here 
\begin{equation}
\begin{split}
H=&\int d^{2}x\Bigg(\frac{1}{2}\Pi^{i}_{(0)}\Pi^{(0)}_{i}+\frac{1}{4}B^{i j}_{(0)}B^{(0)}_{i j}-A^{(0)}_{0}\gamma_{(0)}+\left(-4\partial_{j}B^{j i}_{(0)}+2\partial_{j}F^{j i}_{(0)}\right)\chi^{(0)}_{0 i}+2\chi^{(0)}_{i j}\partial_{j}\Pi^{(0)}_{i}-\tilde{\chi}^{(0)}_{i j}F^{i j}_{(0)}\\
&+\sum_{n=1}^{\mathcal{N}}\Bigg[\frac{1}{2}\Pi^{i}_{(n)}\Pi^{(n)}_{i}+\frac{1}{2}\Pi^{3}_{(n)}\Pi^{(n)}_{3}+\frac{1}{4}B^{i j}_{(n)}B^{(n)}_{i j}+\frac{1}{2}B^{i 3}_{(n)}B^{(n)}_{i 3}-A^{(n)}_{0}\gamma_{(n)}-F^{i j}_{(n)}\tilde{\chi}^{(n)}_{i j}+2\chi^{(n)}_{i j}\partial_{i}\Pi^{(n)}_{j}\\
&-B^{i3}_{(n)}\Big(\partial_{i}A^{(n)}_{3}+\frac{n}{R}A^{(n)}_{i}\Big)+\Big(\partial_{i}A^{(n)}_{3}+\frac{n}{R}A^{(n)}_{i}\Big)\Big(\partial_{i}A^{(n)}_{3}+\frac{n}{R}A^{(n)}_{i}\Big)+2\left(\partial_{i}\Pi^{(n)}_{3}+\frac{n}{R}\Pi^{(n)}_{i}\right)\chi ^{(n)}_{i 3}\\
& +\left(-4\partial_{j}B^{j i}_{(n)}+2\partial_{j}F^{j i}_{(n)}+\frac{2n}{R}\Big(\partial_{i}A^{(n)}_{3}+\frac{n}{R}A^{(n)}_{i}\Big)\right)\chi^{(n)}_{0 i}+\left(-4\partial_{i}B^{i 3}_{(n)}+2\partial_{i}\left(\partial_{i}A^{(n)}_{3}+\frac{n}{R}A^{(n)}_{i}\right)\right)\chi^{(n)}_{0 3}\Bigg]\Bigg)\\
=&\int d^{2}x\Bigg(\mathcal{H}^{(0)}+\sum_{n=1}^{\mathcal{N}}\mathcal{H}^{(n)}\Bigg).
\end{split}
\label{hfcs}
\end{equation}
Note,  that the extended Hamiltonian is not a linear combination of constraints anymore, the term $B^{MN}B_{MN}$ of the action (\ref{ac1}) breaks down the general covariance of the theory and eliminates the reducibility relations  present in the $BF$-like term.\\
Now,  the first class constraints allows us to know the fundamental gauge transformations; for this important step, we  use  the Castellani's  procedure \cite{Castellani, henn} to construct the gauge generators
\begin{eqnarray}
G= \int_\Sigma \left[\varepsilon^{(n)}_{0} \gamma^{0}_{(n)}  +  \varepsilon^{(n)} \gamma_{(n)} +\varepsilon^{(0)}_{0}\gamma^{0}_{(0)} + \varepsilon^{(0)}\gamma_{(0)} \right]dx^2, 
\label{eq62}
\end{eqnarray}
thus, we find that the gauge transformations on the phase space are given for \\
zero modes  
\begin{eqnarray}
\delta A_{\mu}^{(0)}&=&-\partial_{\mu}\varepsilon_{(0)},\nonumber\\
\delta B_{\mu \nu}^{(0)}&=&0,\nonumber\\
\delta \Pi^{\mu}_{(0)}&=& 0,\nonumber\\
\delta \Pi^{\mu \nu}_{(0)}&=& 0,
\label{eq63}
\end{eqnarray}
and the gauge transformation for the  $KK$ modes
\begin{eqnarray}
\delta A_{\mu}^{(n)}&=&-\partial_{\mu}\varepsilon_{(n)},\\ \label{AN}
\delta A_{3}^{(n)}&=&\frac{n}{R}\varepsilon_{(n)},\\ \label{A3}
\delta B_{\mu \nu}^{(n)}&=&0,\\ \label{BT}
\delta \Pi^{\mu}_{(n)}&=& 0,\\
\delta \Pi^{\mu \nu}_{(n)}&=& 0,
\end{eqnarray}
we can observe  that  the gauge transformations for  the zero mode are   the same given for  Maxwell theory written  in the standard form \cite{henn},  and we also observe that the $B$ filed is not a gauge field anymore. Finally,  the transformations of the fields $ A_ {\mu} ^{n} $, $A_3 ^{n}$  corresponding for  the $k$-th mode are the same to those reported in  the literature (see \cite{3, roman} and the cites there in). 
Hence,  by fixing   the gauge parameters   by $\varepsilon_{(n)}=-\frac{R}{n}A_{3}^{(n)}$ and considering the second class constraints as strong identities, the effective action (\ref{eqlaga}) is reduced  to that reported in \cite{3, roman}, namely 
\begin{equation}
{\mathcal{L}}=-\frac{1}{4}F^{\mu \nu}_{(0)}F^{(0)}_{\mu \nu}+\sum_{n=1}^{\mathcal{\infty}}\Bigg[-\frac{1}{4}F^{\nu \mu}_{(n)}F^{(n)}_{\nu \mu}+\frac{1}{2} \left( \frac{2n}{R}\right)^2 A_\mu^{(n)} A^{\mu (n)}\Bigg],
\label{eqef}
\end{equation}
where we able to observe that the KK-modes are massive  Proca fields, and $A_3^{(n)}$ has been absorbed and it is  identified as  a pseudo-Goldstone boson \cite{3, roman}.  Furthermore,  we have added in the   appendix A  the  Dirac  brackets of the theory, thus we have developed a full Hamiltonian analysis of the theory under study.
\section{ Conclusions}
In the context of extra dimensions, the Hamiltonian analysis for a topological $BF$-like theory and  for Maxwell theory expressed as  a  $BF$-like theory  has been performed. For the former,  after  performing the compactification  process  on a $S^1/\mathbf{Z_2}$ orbifold,   we analyzed the effective theory and we have obtained in the canonical analysis all the constraints, gauge transformations  and the extended Hamiltonian. For this theory we found that the extended Hamiltonian is given by  a  linear combination of first class constraints of the zero mode and first class constraints of the KK modes, this indicates that the compactification process does not  break  the general covariance of the theory. Moreover, we observe  that   reducibility relations among the constraints are preserved  before and after  performing the compatification process, however,  after performing  the compactification  the reducibility  is   given  among the firs class constraints of the excited modes, there is not reducibility  in the zero modes. This important fact allowed us to conclude that the theory is  a topological one. \\ 
Finally,  for  Maxwell theory written as a $BF$-like theory with a compact dimension, we found the constraints, the gauge transformations and the extended action. We observed  that the theory does not present reducibility conditions among the constraints,  and the theory  is not topological anymore. In fact, the theory has the same symmetries and  degrees of freedom than Maxwell theory with a compact dimension \cite{3,5}.  Finally by fixing  the gauge parameters we noted  that the theory is reduced to Maxwell theory   in three dimensions described by  the zero mode plus  a tower of massive Proca fields excitations. \\
We would to comment that our results are generic and can be extended to a  5D theory and models with a close  relation to YM and general relativity. In fact, we have commented above that   there are  topological generalizations of  Maxwell and Yang-Mills theories  in three and four  dimensions,  that  could provide generalized QCD theories as it  is  claimed in \cite{8}.  In this manner, our results can be  used for studying   those generalizations in the context of extra dimensions. Furthermore, our results can be used for studying  models  that are present in string theory  such as those models described by a  Kalb-Ramond  field,  or  we can study gravity theories written as a  $BF$  structure \cite{joy}.  However, these  ideas are  still in progress and will be the subject of future  works. \\
\section{Appendix A}
In this section we will compute the Dirac brackets for the $BF$ theory with a compact dimension given by the action (\ref{eq5}). By using the constraints given in (\ref{eq15u}),  (\ref{eq15}) and  the fixed gauge $\partial_{i}A^{(0)}_{i}\approx 0$, $A^{(0)}_{0}\approx 0$, $2\eta_{i j}\partial^{i}B^{0j}_{(0)}\approx 0$ and $2B^{ij}_{(0)}\approx0$ we obtain the  following set of second class constraints 
\begin{eqnarray}
\hat{\chi}^{(0)}&=&\partial_{i}A^{(0)}_{i}\approx 0,\nonumber\\
\chi^{(0)}_{0}&=&A^{(0)}_{0}\approx 0,\nonumber\\
\chi_{(0)}&=&\partial_{i}\Pi^{i}_{(0)} \approx 0,\nonumber \\
\bar{\chi}^{0}_{(0)}&=&\Pi^{0}_{(0)} \approx 0,\nonumber \\
\tilde{\chi}^{(0)}&=&\frac{1}{2}\eta^{i j}F^{(0)}_{ij}-\eta^{i j}\partial_{i}\Pi^{(0)}_{0j}\approx 0,\nonumber \\
\hat{\chi}_{(0)} &=&2\eta^{i j}\partial_{i}B^{0j}_{(0)}\approx 0,\nonumber \\
\chi^{ij}_{(0)}&=&2B^{ij}_{(0)}\approx 0,\nonumber\\
\tilde{\chi}^{(0)}_{ij}&=&\Pi^{(0)}_{ij} \approx 0,\nonumber \\
\chi^{i}_{(0)}&=&\Pi^{i}_{(0)}-2B^{0i}_{(0)}\approx 0,\nonumber\\
\chi^{(0)}_{0i}&=&\Pi^{(0)}_{0i} \approx 0.\nonumber\\
\label{xc}
\end{eqnarray}
Thus, the matrix whose entries are given by the Poisson brackets among the constraints (\ref{xc}) is given by 
{\tiny
\begin{eqnarray}
G^{(0)}_{\alpha\nu}=
\bordermatrix{
 & \hat{\chi}^{(0)} & \chi^{(0)}_{0} & \chi_{(0)} & \bar{\chi}^{0}_{(0)} & \tilde{\chi}^{(0)} & \hat{\chi}_{(0)} & \chi^{kl}_{(0)} &\tilde{\chi}^{(0)}_{kl} &  \chi^{k}_{(0)} & \chi^{(0)}_{0k} \cr
\hat{\chi}^{(0)} & 0 & 0 & -\nabla^{2} & 0 & 0 & 0 & 0 & 0 & \partial_{k} & 0 \cr
\chi^{(0)}_{0} & 0 & 0 & 0 & 1 & 0 & 0 & 0 & 0 & 0 & 0 \cr
\chi_{(0)} &\nabla^{2} & 0 & 0 & 0 & 0 & 0 & 0 & 0 & 0 & 0 \cr
\bar{\chi}^{0}_{(0)} & 0 & -1 & 0 & 0 & 0 & 0 & 0 & 0 & 0 & 0 \cr
\tilde{\chi}^{(0)} & 0 & 0 & 0 & 0 & 0 & -\nabla^{2} & 0 & 0 & 0 & 0 \cr
\hat{\chi}_{(0)} & 0 & 0 & 0 & 0 & \nabla^{2} & 0 & 0 & 0 & 0 & \eta^{ik}\partial_{i} \cr
\chi^{ij}_{(0)}& 0 & 0 & 0 & 0 & 0 & 0 & 0 & \left( \delta{^{i}}_{k} \delta{^{j}}_{l} -\delta{^{i}}_{l} \delta{^{j}}_{k}\right) & 0 & 0 \cr
\tilde{\chi}^{(0)}_{ij} & 0 & 0 & 0 & 0 & 0 & 0 & -\left( \delta{^{i}}_{k} \delta{^{j}}_{l} -\delta{^{i}}_{l} \delta{^{j}}_{k}\right) & 0 & 0 & 0 \cr
\chi^{i}_{(0)} & -\partial_{i} & 0 & 0 & 0 & 0 & 0 & 0 & 0 & 0 & -\delta{^{k}}_{i} \cr
\chi^{(0)}_{0i}& 0 & 0 & 0 & 0 & 0 & -\eta^{ji}\partial_{j} & 0 & 0 & \delta{^{k}}_{i} & 0
 \cr}\delta^{2}(x-y)\nonumber\\
\end{eqnarray}
}
the inverse is given by 
{\tiny
\begin{eqnarray}
G^{(0)-1}_{\alpha\nu}=
\bordermatrix{
& \hat{\chi}^{(0)} & \chi^{(0)}_{0} & \chi_{(0)} & \bar{\chi}^{0}_{(0)} & \tilde{\chi}^{(0)} & \hat{\chi}_{(0)} & \chi^{kl}_{(0)} &\tilde{\chi}^{(0)}_{kl} &  \chi^{k}_{(0)} & \chi^{(0)}_{0k} \cr
\hat{\chi}^{(0)}& 0 & 0 &\frac{1}{\nabla^{2}}& 0 & 0 & 0 & 0 & 0 & 0 & 0 \cr
\chi^{(0)}_{0}& 0 & 0 & 0 & -1  & 0 & 0 & 0 & 0 & 0 & 0 \cr
\chi_{(0)}&-\frac{1}{\nabla^{2}} & 0 & 0 & 0 & 0 & 0 & 0 & 0 & 0 & \frac{\partial_{k}}{\nabla^{2}} \cr
\bar{\chi}^{0}_{(0)}& 0 & 1  & 0 & 0 & 0 & 0 & 0 & 0 & 0 & 0 \cr
\tilde{\chi}^{(0)} & 0 & 0 & 0 & 0 & 0 & \frac{1}{\nabla^{2}} & 0 & 0 & \frac{\eta^{ik}\partial_{i}}{\nabla^{2}} & 0 \cr
\hat{\chi}_{(0)}& 0 & 0 & 0 & 0 & -\frac{1}{\nabla^{2}} & 0 & 0 & 0 & 0 & 0 \cr
\chi^{ij}_{(0)}& 0 & 0 & 0 & 0 & 0 & 0 & 0 & -\left( \delta{^{i}}_{k} \delta{^{j}}_{l} -\delta{^{i}}_{l} \delta{^{j}}_{k}\right) & 0 & 0 \cr
\tilde{\chi}^{(0)}_{ij}& 0 & 0 & 0 & 0 & 0 & 0 & \left( \delta{^{i}}_{k} \delta{^{j}}_{l} -\delta{^{i}}_{l} \delta{^{j}}_{k}\right) & 0 & 0 & 0 \cr
\chi^{i}_{(0)}& 0 & 0 & 0 & 0 & -\frac{\eta^{ji}\partial_{j}}{\nabla^{2}} & 0 & 0 & 0 & 0 & \delta{^{k}}_{i} \cr
\chi^{(0)}_{0i}& 0 & 0 & -\frac{\partial_{i}}{\nabla^{2}}  & 0 & 0 & 0 & 0 & 0 & -\delta{^{k}}_{i} & 0 \cr
 }
\delta^{2}(x-y).
\end{eqnarray}
}
In this manner, the Dirac brackets are given by 
\begin{eqnarray}
\{A^{(0)}_{i}(x), \Pi^{j} _{(0)}(y)\}_{D} &=&\delta^{j}{_{i}}\delta^{2}(x-y)-\frac{1}{{\nabla^{2}}} \left(\partial_{i}\partial_{j}-\eta^{ki}\eta^{lj}\partial_{k}\partial_{l}\right)\delta^{2}(x-y) , \nonumber \\ 
\{B^{0 i}_{(0)}(x), A^{(0)} _{j}(y)\}_{D}&=&-\frac{1}{2}\delta^{j}{_{i}}\delta^{2}(x-y)-\frac{1}{2 \nabla^2}\left(\partial_{i}\partial_{j}-\eta^{ki}\eta^{lj}\partial_{k}\partial_{l}\right) \delta^{2}(x-y), \nonumber \\
\{B^{0 i}_{(0)}(x), \Pi^{(0)} _{0j}(y)\}_{D}&=&0, \nonumber \\
\{B^{i j}_{(n)}(x), \Pi^{(n)} _{kl}(y)\}_{D} &=&0. \nonumber \\
\label{eqbn}
\end{eqnarray}
By means an easy calculation, we can obtain similar results for the excited modes.
\section{ Appendix B}
In this appendix, we calculate the Dirac  brackets for Maxwell theory written as a BF-like theory. For our aims we will calculate the Dirac brackets for the zero mode, then we will calculate the brackets for the excited modes.  Hence,  by using the following fixed gauge $ \partial^i A_i^{(0)} \approx 0$ and $A_{0}^{(0)}\approx 0$, we obtain the following set of second class constraints 
\begin{eqnarray}
\bar{\chi}^{(0)}&=&A^{(0)}_{0}\approx 0,\nonumber \\
\hat{\chi}_{(0)}&=&\Pi^{0}_{(0)}\approx 0,\nonumber \\
\tilde{\chi}^{(0)}&=&\partial_{i}A^{(0)}_{i}\approx 0,\nonumber \\
\chi_{(0)}&=&\partial_{i}\Pi^{i}_{(0)}\approx 0,\nonumber \\
\chi^{(0)}_{0i}&=&\Pi^{(0)}_{0i} \approx 0,\nonumber \\
\chi^{(0)}_{ij}&=&\Pi^{(0)}_{ij} \approx 0,\nonumber \\
\chi^{i}_{(0)}&=&\Pi^{i}_{(0)}+B^{0i}_{(0)} \approx 0,\nonumber \\
\tilde{\chi}^{(0)}_{ij}&=&\frac{1}{2}\left(B^{(0)}_{ij}-F^{(0)}_{ij}\right) \approx 0,\nonumber \\
\end{eqnarray}
thus,  we can calculate  the following matrix whose entries are given by the Poisson brackets  between  these constraints,  obtaining 
{\tiny
\begin{eqnarray}
C^{(0)}_{\alpha\nu}(x,y)=
\bordermatrix{
 & \bar{\chi}^{(0)} & \hat{\chi}_{(0)} & \tilde{\chi}^{(0)} & \chi_{(0)} & \chi^{(0)}_{0 j} &\chi^{(0)}_{kl} & \chi^{k}_{(0)} &\tilde{\chi}^{(0)}_{kl}\cr
\bar{\chi}^{(0)} &0 & 1 & 0 & 0 & 0 & 0 & 0 & 0\cr
\hat{\chi}_{(0)} &-1 & 0 & 0 & 0 & 0 & 0 & 0 & 0\cr
\tilde{\chi}^{(0)} &0 & 0 & 0 &-\nabla^{2} & 0 & 0 & \partial_{k} & 0\cr
\chi_{(0)} &0 & 0 & \nabla^{2} & 0 & 0 & 0 & 0 & 0\cr
\chi^{(0)}_{0 i} &0 & 0 & 0 & 0 & 0 & 0 & -\frac{1}{2}\delta{^{i}}_{k} & 0 \cr
\chi^{(0)}_{ij} &0 & 0 & 0 & 0 & 0 & 0 & 0 & -\frac{1}{4}\left( \delta{^{i}}_{k} \delta{^{j}}_{l} -\delta{^{i}}_{l} \delta{^{j}}_{k}\right)\cr
\chi^{i}_{(0)} &0 & 0 & -\partial_{i} & 0 & \frac{1}{2}\delta{^{i}}_{j} & 0 & 0 & \frac{1}{2}\left( \delta{^{i}}_{l}\partial_{k}-\delta{^{i}}_{k}\partial_{l}\right) \cr
\tilde{\chi}^{(0)}_{i j} &0 & 0 & 0 & 0 & 0 & \frac{1}{4}\left( \delta{^{i}}_{k} \delta{^{j}}_{l} -\delta{^{i}}_{l} \delta{^{j}}_{k}\right) &  -\frac{1}{2}\left( \delta{^{k}}_{j}\partial_{i}-\delta{^{k}}_{i}\partial_{j}\right)& 0\cr}
\delta^{2}(x-y).
\nonumber\\
\end{eqnarray}
}
The inverse of this matrix is given by 
{\tiny
\begin{eqnarray}
C^{(0)-1}_{\alpha\nu}(x,y)=
\bordermatrix{
 & \bar{\chi}^{(0)} & \hat{\chi}_{(0)} & \tilde{\chi}^{(0)} & \chi_{(0)} & \chi^{(0)}_{0 k} &\chi^{(0)}_{k l} & \chi^{k}_{(0)} &\tilde{\chi}^{(0)}_{kl}\cr
\bar{\chi}^{(0)} &0 & -1 & 0 & 0 & 0 & 0 & 0 & 0\cr
\hat{\chi}_{(0)} &1 & 0 & 0 & 0 & 0 & 0 & 0 & 0\cr
\tilde{\chi}^{(0)} &0 & 0 & 0 & \frac{1}{\nabla^{2}} & 0 & 0 & 0 & 0\cr
\chi_{(0)} &0 & 0 & -\frac{1}{\nabla^{2}} & 0 & -\frac{2 \partial_{k}}{\nabla^{2}} & 0 & 0 & 0\cr
\chi^{(0)}_{0 i} &0 & 0 & 0 & \frac{2\partial_{i}}{\nabla^{2}} & 0 & 4\left( \delta{^{i}}_{l}\partial_{k}-\delta{^{i}}_{k}\partial_{l}\right) & 2\delta{^{i}}_{k} & 0 \cr
\chi^{(0)}_{ij} &0 & 0 & 0 & 0 & -4\left( \delta{^{k}}_{j}\partial_{i}-\delta{^{k}}_{i}\partial_{j}\right) & 0 & 0 &4\left( \delta{^{i}}_{k} \delta{^{j}}_{l} -\delta{^{i}}_{l} \delta{^{j}}_{k}\right)\cr
\chi^{i}_{(0)} &0 & 0 & 0 & 0 & -2\delta{^{i}}_{k} & 0 & 0 & 0\cr
\tilde{\chi}^{(0)}_{i j} &0 & 0 & 0 & 0 & 0 &-4\left( \delta{^{i}}_{k} \delta{^{j}}_{l} -\delta{^{i}}_{l} \delta{^{j}}_{k}\right)  & 0& 0\cr}
\delta^{2}(x-y).
\nonumber\\
\end{eqnarray}
}
In this manner, the Dirac brackets of two functionals $A$, $B$ defined on the phase space,  is expressed by
\[
\{F(x),G(z)\}_{D}\equiv\{F(x),G(z)\}-\int d^{2}ud^{2}w\{F(x),\xi_{\alpha}(u)\}C{^{\alpha\beta}}\{\xi_{\beta}(w),G(z)\},
\]
where $\{F(x),G(z)\}$ is the Poisson bracket  between two functionals $F,G$,  and $\xi_{\alpha}$ represent  the set of second class constraints. By using this fact, we obtain the following  Dirac's brackets for the zero mode 
\begin{eqnarray}
\{A^{(0)}_{i}(x), \Pi^{j} _{(0)}(y)\}_{D} &=& \left( \delta^{j}{_{i}} - \frac{\partial _{j} \partial_{i}}{ \nabla ^{2}} \right) \delta^{2} (x-y), \nonumber \\ 
\{B^{i j}_{(0)}(x), \Pi^{(0)} _{kl}(y)\}_{D} &=&0 \nonumber \\
\{A^{(0)}_{i }(x), \Pi^{j}_{(0)}(y)\}_{D} &=&0 \nonumber \\
\{A^{(0)}_{i }(x), A^{(0)}_{j}(y))\}_{D} &=&0 \nonumber \\
\{\Pi^{i }_{(0)}(x), \Pi^{j}_{(0)}(y)\}_{D} &=&0 \nonumber \\
\{B^{i j}_{(0)}(x), \Pi^{k} _{(0)}(y)\}_{D} &=&2\left(\delta^{k}{_{j}}\partial_{i}-\delta^{k}{_{i}}\partial_{j}\right)\delta^{2}(x-y),\nonumber \\
\{B^{i j}_{(0)}(x), B^{0 k}_{(0)}(x)\}_{D} &=&-2\left(\delta^{k}{_{j}}\partial_{i}-\delta^{k}{_{i}}\partial_{j}\right)\delta^{2}(x-y),\nonumber \\
\{A^{(0)}_{i}(x), B^{0 j}_{(0)}(x)\}_{D} &=&-\left( \delta^{j}{_{i}} - \frac{\partial _{j} \partial_{i}}{ \nabla ^{2}} \right) \delta^{2} (x-y),
\label{eqbn}
\end{eqnarray}
we can observe that the Dirac brackets among the fields $A_i^{(0)}, \Pi^j_{(0)}$ are those knew for Maxwell theory \cite{henn}. \\
Now we calculate Dirac's brackets for the excited modes of the Maxwell  $BF$-like theory. By working with the following fixed gauge $\partial_{i}A^{(n)}_{i}\approx 0$ and $ \Pi^{3}_{(n)}+\frac{n}{R}A^{(n)}_{0}\approx0 $, 
we obtain the following set of second class constraints 
\begin{eqnarray}
\tilde{\chi}^{(n)}&=&\partial_{i}A^{(n)}_{i}\approx 0,\nonumber \\
\chi^{0}_{(n)}&=&\Pi^{0}_{(n)}\approx 0,\nonumber \\
\tilde{\chi}^{3}_{(n)}&=&\Pi^{3}_{(n)}+\frac{n}{R}A^{(n)}_{0}\approx 0,\nonumber \\
\chi_{(n)}&=&\partial_{i}\Pi^{i}_{(n)}+\frac{n}{R}\Pi^{3}_{(n)}\approx 0,\nonumber \\
\chi^{(n)}_{0j} &=&\Pi^{(n)}_{0j} \approx 0,  \nonumber \\
\chi^{(n)}_{ij} &=&\Pi^{(n)}_{ij} \approx 0,  \nonumber \\
\chi^{i}_{(n)}&=&\Pi^{i}_{(n)}+B^{0i}_{(n)} \approx 0, \nonumber \\
\tilde{\chi}^{(n)}_{ij}&=&\frac{1}{2}\left(B^{(n)}_{ij}-F^{(n)}_{ij}\right) \approx 0,\nonumber\\
\chi^{(n)}_{03} &=&\Pi^{(n)}_{03} \approx 0,  \nonumber \\
\chi^{(n)}_{i3} &=&\Pi^{(n)}_{i3} \approx 0,  \nonumber \\
\chi^{3}_{(n)}&=&\Pi^{3}_{(n)}+B^{03}_{(n)} \approx 0, \nonumber \\
\tilde{\chi}^{(n)}_{i3}&=&\frac{1}{2}\left(B^{(n)}_{i3}-\left(\partial_{i}A^{(n)}_{3}+\frac{n}{R}A^{(n)}_{i}\right)\right) \approx 0,  \nonumber \\  
\label{eqrg}
\end{eqnarray}
thus,  we obtain the following matrix whose entries are given by the Poisson brackets  between  these second class constraints,  obtaining 
{\tiny
\begin{eqnarray}
G^{(n)}_{\alpha\nu}=
\bordermatrix{
 & \tilde{\chi}^{(n)} & \chi^{0}_{(n)} & \tilde{\chi}^{3}_{(n)} & \chi_{(n)} & \chi^{(n)}_{0 k} &\chi^{(n)}_{k l} & \chi^{k}_{(n)} &\tilde{\chi}^{(n)}_{kl}&\chi^{(n)}_{03}&\chi^{(n)}_{k3}&\chi^{3}_{(n)} &\tilde{\chi}^{(n)}_{k3}\cr
\tilde{\chi}^{(n)} & 0 & 0  & 0  & -\nabla^{2}  & 0  & 0  & \partial_{k}  &  0 & 0  & 0  & 0  & 0  \cr
\chi^{0}_{(n)} & 0  & 0  & -\frac{n}{R}  & 0  & 0  & 0  & 0  & 0  & 0  & 0  & 0  & 0  \cr
\tilde{\chi}^{3}_{(n)} & 0  & \frac{n}{R} & 0  & 0  & 0  & 0  & 0  & 0  & 0  & 0  & 0  & \frac{1}{2}\partial_{k}  \cr
\chi_{(n)} & \nabla^{2}  & 0  & 0  & 0  & 0  & 0  & 0  & 0  & 0  & 0  & 0  & 0  \cr
\chi^{(n)}_{0 i} & 0  & 0  & 0  & 0  & 0  & 0  & -\frac{1}{2}\delta{^{i}}_{k}   & 0  & 0  & 0  & 0  & 0  \cr
\chi^{(n)}_{i j} & 0  & 0  & 0  & 0  & 0  & 0  & 0  & -\frac{1}{4}\left( \delta{^{i}}_{k} \delta{^{j}}_{l} -\delta{^{i}}_{l} \delta{^{j}}_{k}\right)  & 0  & 0  & 0  & 0  \cr
\chi^{i}_{(n)} &- \partial_{i}  & 0  & 0  & 0  &  \frac{1}{2}\delta{^{i}}_{k}  & 0  & 0  & \frac{1}{2}\left( \delta{^{i}}_{l}\partial_{k}-\delta{^{i}}_{k}\partial_{l}\right)  & 0  & 0  & 0  & \frac{n}{2R}\delta{^{i}}_{k}\cr
\tilde{\chi}^{(n)}_{i j} & 0  & 0  & 0  & 0  & 0  & \frac{1}{4}\left( \delta{^{i}}_{k} \delta{^{j}}_{l} -\delta{^{i}}_{l} \delta{^{j}}_{k}\right)  &  -\frac{1}{2}\left(  \delta{^{k}}_{j}\partial_{i}-\delta{^{k}}_{i}\partial_{j}\right)  & 0  & 0  & 0  & 0  & 0 \cr
\chi^{(n)}_{03} & 0  & 0  & 0  & 0  & 0  & 0  & 0  & 0  & 0  & 0 & -\frac{1}{2}  & 0  \cr
\chi^{(n)}_{i3} & 0  & 0  & 0  & 0  & 0  & 0  & 0  & 0  & 0  & 0  & 0  & -\frac{1}{4}\delta{^{i}}_{k}  \cr
\chi^{3}_{(n)} & 0  & 0  & 0  & 0  & 0  & 0  & 0  & 0  & \frac{1}{2}  & 0  & 0  & \frac{1}{2}\partial_{k}\cr
\tilde{\chi}^{(n)}_{i 3} & 0  &  0 & -\frac{1}{2}\partial_{i}  & 0  & 0  & 0 & -\frac{n}{2R}\delta{^{i}}_{k}  & 0  & 0  & \frac{1}{4}\delta{^{i}}_{k}   & -\frac{1}{2}\partial_{i}  & 0\cr
} 
\delta^{2}(x-y)
\nonumber\\
\end{eqnarray}
}
hence, the inverse matrix is given by 
{\tiny
\begin{eqnarray}
G_{\alpha\nu}^{(n)-1}=
\bordermatrix{
 & \tilde{\chi}^{(n)} & \chi^{0}_{(n)} & \tilde{\chi}^{3}_{(n)} & \chi_{(n)} & \chi^{(n)}_{0 k} &\chi^{(n)}_{k l} & \chi^{k}_{(n)} &\tilde{\chi}^{(n)}_{kl}&\chi^{(n)}_{03}&\chi^{(n)}_{k3}&\chi^{3}_{(n)} &\tilde{\chi}^{(n)}_{k3}\cr
\tilde{\chi}^{(n)}& 0  & 0  & 0  & \frac{1}{\nabla^{2}}  & 0  & 0  & 0  & 0  & 0  & 0  & 0  & 0  \cr
\chi^{0}_{(n)} & 0  & 0  & \frac{R}{n}  & 0  & 0  & 0  & 0  & 0  & 0  & \frac{2R}{n}\partial_{k}  & 0  & 0  \cr
\tilde{\chi}^{3}_{(n)}& 0  & -\frac{R}{n}  & 0  & 0  & 0  & 0  & 0  & 0  & 0  & 0  & 0  & 0  \cr
\chi_{(n)}&-\frac{1}{\nabla^{2}}  & 0  & 0  & 0  & -2\frac{\partial_{k}}{\nabla^{2}}  & 0  & 0  & 0  & 0  & 0  & 0  & 0  \cr
\chi^{(n)}_{0 i}&0  & 0  & 0  & 2\frac{\partial_{k}}{\nabla^{2}}  & 0  & 4\left( \delta{^{i}}_{l}\partial_{k}-\delta{^{i}}_{k}\partial_{l}\right)  & 2\delta{^{i}}_{k}  & 0  & 0  & \frac{4n\delta{^{i}}_{k}}{R}  & 0  & 0  \cr
\chi^{(n)}_{ij}& 0  & 0  & 0  & 0  & -4\left( \delta{^{k}}_{j}\partial_{i}-\delta{^{k}}_{i}\partial_{j}\right)  & 0  & 0  & 4\left( \delta{^{i}}_{k} \delta{^{j}}_{l} -\delta{^{i}}_{l} \delta{^{j}}_{k}\right)  & 0  & 0  & 0  & 0  \cr
\chi^{i}_{(n)}& 0  & 0  & 0  & 0  & -2\delta{^{i}}_{k}  & 0  & 0  & 0  & 0  & 0  & 0  & 0  \cr
\tilde{\chi}^{(n)}_{ij}& 0  & 0  & 0  & 0  & 0  & -4\left( \delta{^{i}}_{k} \delta{^{j}}_{l} -\delta{^{i}}_{l} \delta{^{j}}_{k}\right) & 0  & 0  & 0  & 0  & 0  & 0  \cr
\chi^{(n)}_{03}& 0  & 0  & 0  & 0  & 0  & 0  & 0  & 0  & 0  & 4\partial_{k}  & 2  & 0  \cr
\chi^{(n)}_{i3}& 0  & -\frac{2R}{n}\partial_{i}  & 0  & 0  & -\frac{4n\delta{^{i}}_{k}}{R} & 0  & 0  & 0  & -4\partial_{i}  & 0  & 0  & 4\delta{^{i}}_{k}  \cr
\chi^{3}_{(n)} & 0  & 0  & 0  & 0  & 0  & 0  & 0  & 0  & -2  & 0  & 0  & 0  \cr
\tilde{\chi}^{(n)}_{i3}& 0  & 0  & 0  & 0  & 0  & 0  & 0  & 0  & 0  & -4\delta{^{i}}_{k}  & 0  & 0 \cr
 }
\delta^{2}(x-y).
\nonumber\\
\end{eqnarray}
}
In this manner, the Dirac brackets for the excited modes are given by 
\begin{eqnarray}
\{A^{(n)}_{i}(x), \Pi^{j}_{(n)} (y)\}_{D} &=& \bigg(\delta^{j}{_{i}}-\frac{\partial_{i}\partial_{j}}{\nabla^{2}}\bigg)\delta^{2}(x-y),\nonumber \\ 
\{A_{3}^{(n)} (x),\Pi^{i} _{(n)}(y)\}_{D} &=&\frac{n}{R}\partial_{i}\left(\frac{\delta^{2}(x-y)}{\nabla^{2}}\right)\nonumber\\
\{A^{(n)}_{3}(x), \Pi^{3} _{(n)}(y)\}_{D} &=&\delta^{2}(x-y), \nonumber \\
\{B^{i j}_{(n)}(x), \Pi^{(n)} _{kl}(y)\}_{D} &=&0, \nonumber \\
\{B^{i j}_{(n)}(x), \Pi^{k} _{(n)}(y)\}_{D} &=&2\left(\delta^{k}{_{j}}\partial_{i}-\delta^{k}{_{i}}\partial_{j}\right)\delta^{2}(x-y),\nonumber \\
\{B^{i j}_{(n)}(x), B^{0 k}_{(n)}(x)\}_{D} &=&-2\left(\delta^{k}{_{j}}\partial_{i}-\delta^{k}{_{i}}\partial_{j}\right)\delta^{2}(x-y),\nonumber \\
\{B^{i 3}_{(n)}(x), B^{0 3}_{(n)}(x)\}_{D} &=&-\partial_{i}\delta^{2}(x-y),\nonumber \\
\{B^{i3}_{(n)}(x), \Pi^{j} _{(n)}(y)\}_{D} &=&\frac{n}{R}\delta^{i}{_{j}}\delta^{2}(x-y), \nonumber \\
\{B^{i3}_{(n)}(x), \Pi^{3} _{(n)}(y)\}_{D} &=&\partial_{i}\delta^{2}(x-y), \nonumber \\ 
\{A^{(n)}_{i}(x), B^{0 j}_{(n)}(x)\}_{D} &=&-\left( \delta^{j}{_{i}} - \frac{\partial _{j} \partial_{i}}{ \nabla ^{2}} \right) \delta^{2} (x-y),\nonumber \\
\{A^{(n)}_{3}(x), B^{0 3}_{(n)}(x)\}_{D} &=&\delta^{2} (x-y).\nonumber \\
\label{dbmebf}
\end{eqnarray}
Therefore, we have in this work all the elements for  studying  the quantization of the theories under study. 
\newline
  \noindent \textbf{Acknowledgements}\\[1ex]
  This work was supported by the Sistema Nacional de Investigadores (M\'exico). We would to thank  R. Cartas-Fuentevilla for reading the manuscript. 


\begin{thebibliography}{}
\setlength{\itemsep}{-.50em}
\bibitem{1} Th. Kaluza, Sitzungober. Preuss. Akad. Wiss. Berlin (1921) 966;
O. Klein, Z. Phys. 37 (1926) 895.
\bibitem{1a} Merab Gogberashvili,  Alfredo Herrera Aguilar, Dagoberto
Malag\'on Morej\'on and Refugio Rigel Mora Luna, Phys. Lett. B 725, 208-211,  (2013),  arXiv:1202.1608.
\bibitem{2} M. B. Green, J. H. Schwarz and E. Witten, Superstring Theory (Cambridge University
Press, Cambridge, 1986); J. Polchinski, String Theory (Cambridge University Press, Cambridge, 1998);  S. T. Yau (ed.), Mathemathical Aspects of String Theory (World Scientific, Singapore, 17 1987).
\bibitem{3} A. P\'erez-Lorenzana, J. Phys. Conf. Ser. 18, 224 (2005).
 \bibitem{6} A. Muck, A. Pilaftsis and R. Ruckl, Phys. Rev. D 65, 085037 (2002).
 \bibitem{6a} I. Antoniadis, Phys. Lett. B 246, 377, (1990). 
\bibitem{6b}  J.D. Lykken, Phys. Rev. D 54, 3693 (1996).
\bibitem{6d} K.R. Dienes, E. Dudas, and T. Gherghetta, Phys. Lett. B 436, 55 (1998); Nucl. Phys. B537, 47 (1999).
 \bibitem{5}  L. Nilse, hep-ph/0601015. 
\bibitem{LHC-ILC}  G. Weiglein \textit{et al.} (Physics Interplay of the LHC and the ILC), \href{http://xxx.lanl.gov/abs/hep-ph/0410364}{arXiv:hep-ph/0410364}.
\bibitem{7} H. Novales-Sa«nchez and J. J. Toscano, Phys. Rev. D, 82, 116012 (2010).
\bibitem{7a}  L. Freidel and A. Starodubtsev, Quantum gravity in terms of topological observables,
preprint (2005), arXiv: hep-th/0501191.
\bibitem{7b} A. Escalante, Phys. Lett. B 676 (2009) 105Ð111.
\bibitem{8} J. Govaerts, { \it Topological quantum field theory and pure YangÐMills dynamics}, in
Proc. Third Int. Workshop on Contemporary Problems in Mathematical Physics
(COPROMAPH3 ), Cotonou (Republic of Benin), 1Ð7 November 2003.
\bibitem{9}A. Escalante, J. Lopez-Osio, Int.J.Pure Appl.Math. \textbf{75} 339-352, (2012). 
\bibitem{sund} Sundermeyer K. { \it Constrained Dynamics}, Lecture Notes in Physics vol.169, Spinger-Verlag, Berlin Heidelberg New York, 1982.
\bibitem{10} A. Escalante, J. Berra, Int.J.Pure Appl.Math. \textbf{79} 405-423, (2012). 
\bibitem{12} M. Mondragon, M. Montesinos, J. Math. Phys., 47, 022301, (2006).
\bibitem{Muck:2002af}A.~Muck, A.~Pilaftsis and R.~Ruckl, Lect.\ Notes Phys.\  {\bf 647}, 189 (2004).
\bibitem{11} A. Escalante and  Ira{\'i}s Rubalcava, Int. J. Geom. Meth. Mod. Phys Vol. 9, No. 7, 1250053, (2012).
\bibitem{Castellani} L. Castellani, Annals Phys. \textbf{143}, 357 (1982).
\bibitem{henn} Henneaux M.~and Teitelboim C. { \it Quantization of Gauge Systems}, Princeton University Press, Princeton, New Jersey, 1992.
\bibitem{roman}  G. De los Santos and R. Linares, AIP Conf. Proc. \textbf{1256}, 178 (2010).
\bibitem{joy} Alberto Escalante and Mois\'es Z\'arate Reyes, {\it Study of  gravity and string models  in the context of extra dimensions}, in preparation, (2014).


\end{thebibliography}
\end{document}